\documentclass[sigconf]{acmart}

\AtBeginDocument{%
  \providecommand\BibTeX{{%
    \normalfont B\kern-0.5em{\scshape i\kern-0.25em b}\kern-0.8em\TeX}}}

\copyrightyear{2024}
\acmYear{2024}
\setcopyright{acmlicensed}\acmConference[ESEM '24]{Proceedings of the 18th ACM / IEEE International Symposium on Empirical Software Engineering and Measurement}{October 24--25, 2024}{Barcelona, Spain}
\acmBooktitle{Proceedings of the 18th ACM / IEEE International Symposium on Empirical Software Engineering and Measurement (ESEM '24), October 24--25, 2024, Barcelona, Spain}
\acmDOI{XXXXXX/XXXXXXXXXXXXX}
\acmISBN{XXXXXXXXXXXXXXXX}

\usepackage{graphicx}
\graphicspath{ {./images/} }
\usepackage[export]{adjustbox}
\usepackage{array}
\usepackage{xcolor}
\usepackage{listings}
\definecolor{light-gray}{gray}{0.95}
\begin{document}

\title[Debugging with Open-Source Large Language Models: An Evaluation]{Debugging with Open-Source Large Language Models:\\ An Evaluation}


\author {Yacine Majdoub}
\affiliation{
  \institution{University of Gabes, Tunisia}
  \country{}
}
\email{yacinemajdoub@fsg.u-gabes.tn}

\author {Eya Ben Charrada}
\affiliation{
  \institution{University of Gabes, Tunisia}
  \country{}
}
\email{eya.bencharrada@fsg.rnu.tn}

\renewcommand{\shortauthors}{---}


\begin{abstract}

Large language models have shown good potential in supporting software development tasks. This is why more and more developers turn to LLMs (e.g. ChatGPT) to support them in fixing their buggy code. While this can save time and effort, many companies prohibit it due to strict code sharing policies. To address this, companies can run open-source LLMs locally. But until now there is not much research evaluating the performance of open-source large language models in debugging. This work is a preliminary evaluation of the capabilities of open-source LLMs in fixing buggy code. The evaluation covers five open-source large language models and uses the benchmark DebugBench which includes more than 4000 buggy code instances written in Python, Java and C++. Open-source LLMs achieved scores ranging from 43.9\% to 66.6\% with DeepSeek-Coder achieving the best score for all three programming languages.  

\end{abstract}

\keywords{Debugging, Large Language Models, Open-Source LLMs}

\maketitle

\section{Introduction}

 \vspace{2mm}

 \textit{"I'd spend an hour figuring out what exactly goes wrong, then five minutes writing the code to fix it, and then half an hour testing the whole thing. That's just over 5\% coding vs. almost 95\% non-coding."}\footnote{text taken from an answer on stackoverflow regarding the time spent debugging https://softwareengineering.stackexchange.com/a/93323} 

 \vspace{2mm}
 
Debugging is known to be time consuming and frustrating. Therefore it is not surprising to find out that developers are turning to large language models to help them solve their problems. In a study with practitioners, Khojah et al.~\cite{khojah2024beyond} found that software engineers were found to turn often to chatGPT for assistance in various software engineering tasks.

Recent research showed promising results in using LLMs for software Engineering tasks in general and for debugging in particular.  For example, LLMs were able to perform well in bug reproduction~\cite{kang2023large}, fault localisation~\cite{wu2023large} and program repair~\cite{xia2023automated}. 
Despite these advantages, using current state of the art LLMs such as 
ChatGPT can be inappropriate for practitioners due to code sharing policies. In fact, most companies consider their code to be private and don't want it to be sent to LLMs run by third parties.
A solution to this problem would be to run an open source LLM locally. 
So far, there has been very limited assessments of the debugging capabilities of open-source large language models. In fact, earlier works mostly focus on evaluating code generation capabilities, for which many benchmarks exist such as the famous OpenAI's HumanEval~\cite{chen2021evaluating} and its descendants (e.g. HumanEval+~\cite{liu2024your} and  Multilingual HumanEval~\cite{ath2022multi}) or the Google's MBPP~\cite{austin2021program}. 

The goal of this work is to evaluate and compare the capabilities of  open-source  large language models in performing debugging tasks. We would like to answer the following two research questions:
\begin{itemize}
    \item RQ1: How do open source LLMs perform in debugging? To answer this question, we use benchmarking to evaluate five open-source LLMs. The benchmark we used includes more than 4000 buggy code instances in Python, C++ and Java.
    \item RQ2: How does the performance of open-source LLMs in code generation impact their performance in debugging? We compare the scores that the LLMs obtained for debugging with the scores that they achieved for coding as evaluated by the HumanEval Benchmark.
\end{itemize}

Our evaluation suggests that although less capable than the most advanced closed source models (e.g. GPT-4), some open source models were able to achieve decent results compared to their relatively small size. For instance, DeepSeek-coder-instruct, which has only 34B parameters, achieved a score above 63\% in all three programming languages. We also found that except for DeepSeek-coder, all models that achieved a higher scores in HumanEval also got better scores in debugging. 

The contributions of this work are:
\begin{itemize}
    \item We conduct an empirical study that evaluates the debugging capabilities of  open source Large Language Models using a large benchmark that includes a few thousands of buggy code instances
    \item We compare the debugging capabilities of the open source LLMs to their coding capabilities as evaluated by the HumanEval benchmark
    \item We provide an extensive discussion of the strengths and limitations of current debugging and coding benchmarks
\end{itemize}


\section{Open Source Large Language Models} \label{models}
There are many open-source LLMs available in the market. Although nearly\footnote{All models we found used the transformer architecture.} all models use the transformer architecture, they differ in their capabilities due to various factors such as model size, quality and volume of training data, and fine-tuning methods.

For this evaluation, we selected five reputed models. Four of them are code models, while the last one is a general-purpose model.

\subsection{Code models}
We selected the coding models that  achieved the best results on the HumanEval benchmark~\cite{chen2021evaluating}. HumanEval is a code generation benchmark released by OpenAI that includes 146 coding tasks.
We present each of the coding models in the following paragraphs.

\subsubsection{Code Llama} Code Llama~\cite{roziere2024code}
is a  family of large language models that is specialised for code, based on Llama2. Code Llama models have been created by fine-tuning the general language model Llama2 using code specific datasets. The developers of Codellama found that for a given budget, fine-tuning the generic Llama2 to generate code outperforms the same architecture trained on code only.
The training was done with publicly available code (mostly near-deduplicated dataset), which includes 8\% of natural language text related to code such as discussions or questions and answers including code snippets. In addition to supporting several natural languages, the Code Llama models are trained to handle long contexts of up to 100K tokens. 
Meta AI released Codellama in three main variants namely (1) \textit{Code Llama}, which is the foundation model (2) \textit{Code Llama - Python}, which is specialized for python code generation and \textit{Code Llama - Instruct}, which is fine-tuned to follow human instructions. All models are available in four sizes: 7B, 13B, 34B and 70B.

For this evaluation, we use the Code Llama - Instruct 70B variant. This variant was trained using 1 trillion tokens and achieved the best performance on HumanEval with a 67.8\% pass@1.

\subsubsection{Phind-Codellama}
Phind-Codellama~\cite{phind} is a fine-tuned version of Code Llama 34B. The first version of Phind-Codellama was fine-tuned on a dataset of nearly 80,000 programming problems and their corresponding solutions. The second version is Phind-CodeLlama-34-v2, which was initialised from the first version, was trained on 1.5B additional tokens. Although Phind-Codellama has smaller number of parameters  compared to the larger Code Llama 70B, it was able to achive relatively high results on HumanEval. For instance Phind-CodeLlama-34B-v2 achieved 73.8\% pass@1 on HumanEval.

\subsubsection{WizardCoder} WizardCoder~\cite{luo2023wizardcoder} is a family of LLMs that use the Evol-Instruct method~\cite{xu2023wizardlm}, an instruction fine tuning method that  makes the code instructions more complex and which enhances the performance of  coding models. 
Wizardcoder is available in five different sizes ranging from 1B to 33B parameters. The 15B version of WizardCoder~\cite{luo2023wizardcoder}, the results of a collaboration between researchers from Microsoft and researchers from Haong Kong Baptist University, is a fine-tuned version of StarCoder~\cite{li2023starcoder} and it achieved 57.3 \%  pass@1 on HumanEval. The 33B version is  trained from the DeepSeek-Coder-base model and achieved 79.9\%  pass@1 on HumanEval~\cite{chen2021evaluating}.   In this evaluation we use the WizardCoder-33B-V1.1.

\subsubsection{Deepseek-Coder} DeepSeek-Coder~\cite{deepseek-coder} is a series of code models trained on a dataset comprising 2 trillion tokens from 87 programming languages. The dataset  is composed of  87\% code and 13\% natural language in English and Chinese. The model is available in various sizes, from 1.3B to 33B parameters.
The DeepSeek-Coder-Instruct variant is an enhancement of the base model that was fine-tuned with an additional 2 billion tokens of instruction data. This improved the model's ability to execute coding tasks given using human instructions. DeepSeek-Coder-Base 33B achieved 50.3\%  pass@1 on HumanEval, while  DeepSeek-Coder-Instruct-33B achieved 69.2\%  pass@1 on HumanEval. We used DeepSeek-Coder-Instruct-33B in our evaluation.  

\subsection{General-Purpose model: Llama 3}

The last model we chose is Llama3, a general purpose LLM. We selected it because it is the best open source LLM available for now, and we wanted to compare its capabilities to the code-specialized large language models.

LLama3, which is developed by Meta AI,  was released in two sizes: 8B and 70B each with a pre-trained and instruction finetuned version. 
Data quality was a major focus for LLama 3, the model has been pre-trained on over 15 trillion high-quality tokens from publicly available sources, seven times more than LLama 2. The training data incorporates four times more coding data to boost capabilities in that domain and over 5\% of the data covers 30+ languages beyond English. The dataset was filtered using a serie of filtering pipelines, heuristic filtering, NSFW detection, deduplication, and quality classifiers.
The model also utilizes a more efficient tokenizer compared to the previous models of Meta AI, and it uses grouped query attention (GQA) to improve inference efficiency and to handle sequences of up to 8,192 tokens.

Llama3 8B achieved 62.2\% pass@1 on HumanEval while Llama3 70B  achieved 81.7\% pass@1~\cite{llama3}. For this evaluation, we used Llama3 70B. 

\section{Study Design}\label{studyDesign}
The evaluation is done with Benchmarking. Benchmarking can be used to efficiently compare different methods, techniques and tools in empirical software engineering \cite{hasselbring2021benchmarking}\cite{acmstandard} .

 We have chosen the benchmark DebugBench \cite{tian2024debugbench}, one of the largest and most recent benchmarks for debugging.
In this section, we present the benchmark and the experimental setup used for the evaluation.

\subsection{Benchmark}


 DebugBench~\cite{tian2024debugbench} is a benchmark designed to evaluate the debugging capabilities of Large Language Models. It consists of a dataset of 4,253 instances of buggy code, collected from code solutions in LeetCode. The goal of DebugBench is to provide a larger scale evaluation that covers fine-grained bug types, and mitigates data leakage risks. 

DebugBench has two main advantages:  (1) it uses code problems that are quite challenging not only for developers but also for large language models and (2) it provides a comprehensive test suit that allows verifying whether the bug was fixed or not. 

\subsubsection{Data}
Tian et al~\cite{tian2024debugbench} created the dataset using problem descriptions and code solutions from LeetCode. The authors used GPT-4 to automatically introduce bugs to the code and then used human inspections to check the integrity of the benchmark. 


To minimize the risk of leakage the authors used code that was released on LeetCode after June 2022  with the average release date being April 2023. To ensure that the extracted code is correct, the authors selected only code that passes all the tests related to it.

The authors of the benchmark develop a bug taxonomy based on Barr's classification criteria that covers four major bug categories (Syntax, Reference, Logic and Multiple) as well as 18 minor bug types. The bugs were introduced by instructing GPT-4 to add a certain type of bugs to the code. Since GPT sometimes fails in including bugs in the code, the authors filtered out the code that does not fail certain tests. The benchmark includes 761 instances with syntax errors, 684 instances with reference errors, 590 instances with logic errors and 2218 instances with multiple errors. A description of the number of instances for each programming language is provided in Table~\ref{tab:NumbInst}

\begin{table}
\caption{Number of buggy code instances per programming language in DebugBench }
\label{tab:NumbInst}
\begin{tabular}{c|c}
\hline
\textbf{Programming Language}              & \textbf{Buggy Instances} \\ \hline
C++      & 1438       \\ 
Java      & 1401   \\ 
Python    & 1414  \\ 
Total    & 4253  \\ \hline
\end{tabular}
\vspace*{-0.5cm}
\end{table}

\subsubsection{Metrics}
DebugBench assesses whether  a bug is fixed or not by using a set of tests that is provided by LeetCode. If all tests pass, then the bug is considered to be fixed, otherwise, the bug is not fixed. The metric used in DebugBench is the \textit{Pass Rate}, which is the number of bugs for which all corresponding tests have passed (repaired bugs) divided by the total number of bugs.~\cite{tian2024debugbench}. More formally, the Pass Rate is defined as follows:  for each buggy code $\theta_i$ and its fixed version $\theta^{*}_i$, 
there is a set of test cases: $(x^{0}_i, y^{0}_i )$, $(x^{1}_i, y^{1}_i )$, ..., $(x^{m}_i , y^{m}_i)$ to test it, where $x_i$ is the input and $y_i$ is the corresponding desired output. Let $a_\theta(x)=y$ denote a program a, based on the script $\theta$ that maps input $x$ to output $y$ .
A bug is considered to be repaired if all tests pass  which can be referred to as \[
\bigwedge_{j=0}^{m} [a_{\theta^{*}_i}(x_i^j) = y_i^j]
\]

This criteria allows for a conservative measure of the bug fixing capabilities. 
In fact, if the instance contains multiple bugs, it is considered to be repaired only if \textbf{all} bugs within the instance are fixed. For $n$ bug instances, the pass rate would be:

\[
PR = \sum_{i=0}^{n} \frac{\bigwedge_{j=0}^{m_{j}} [a_{\theta^{*}_{i}}(x_{i}^{j}) = y_{i}^{j}]}{n} \times 100\%
\]

\subsection{Experimental setup}

To run the open-source LLMs, we use the platform together.ai\footnote{https://www.together.ai/}. This platform offers an interface that we used to prompt each of the models and get their responses. To test whether the fix is correct or not we use the LeetCode test suite to test the proposed answer. The tests were accessed using an interface to LeetCode.   If all tests passed, the fix was considered as successful.

\paragraph{Prompt design} We used a zero-shot prompting.  The prompt  given to the LLM contained a description of the problem to be solved, the buggy code, as well as a description of the task to be done, i.e. debugging the code. We tweaked the prompts used by the original DebugBench paper so that we only get the fixed code in the output. The prompt we used is presented in Listing~\ref{prompt}.

\lstset{basicstyle=\ttfamily\footnotesize,
    backgroundcolor=\color{light-gray}, xleftmargin=0cm,
    frame=tlbr, framesep=0.2cm, framerule=0pt,
}
\begin{lstlisting}[
frame=single, caption={Prompt},label=prompt,breaklines=true]
Observe the following <<Programming language>>
faulty code which is complete with no extra context. 
Your task is to fix up the code, not change it.
You have to write the fixed code (class Solution). 
Do not write anything else beside the fixed code 
in your response.
Buggy Code : <<Buggy code>>
Goal : <<Description>>
\end{lstlisting}

\paragraph{Code extraction}
Some models returned additional explanatory text with the code, as shown  in the WizardCoder example in Listing~\ref{output_example}. To avoid failing the tests because of the additional text, we wrote a script to extract the code (class Solution) from the  answer generated by the LLM.  The extracted code is then used to run the automated tests.
\vspace*{-0.05cm}
\begin{lstlisting}[
frame=single, caption={An output example of WizardCoder}, label=output_example, breaklines=true]
Solution:
class Solution:
    def findComplement(self, num: int) -> int:
        com = ''
        while num > 0 :
            if num % 2 == 1:
                com = '0' + com
            else:
                com = '1' + com
            num = num // 2
        return int(com,2)
Explanation:
The bug in the code is that the com string is not being 
updated properly. In the if block, we are adding 0 
instead of '0' and in the else block, we are adding 1 
instead of '1'. We need to add the string representation
of 0 and 1 to the com string. Also, we need to return the 
integer value of the binary string, so we need to remove 
the first character of the string before converting it to 
integer.
\end{lstlisting}

\begin{table*}[ht]
\caption{Debugging performance \& running costs of the evaluated open-source large language models}
\label{tab:results_table}
\begin{adjustbox}{max width=1\textwidth,center}
\renewcommand{\arraystretch}{1.5}%
\centering
\begin{tabular}{l|ll|ll|ll|l|l}
\hline
\multicolumn{1}{l|}{\textbf{Model}} & \multicolumn{2}{c|}{\textbf{Python}}                              & \multicolumn{2}{c|}{\textbf{Java}}                                & \multicolumn{2}{c|}{\textbf{C++}}                                 & \multicolumn{1}{c|}{\textbf{Final score}} & \multicolumn{1}{c}{\textbf{Costs}} \\ \cline{2-7}
\multicolumn{1}{c|}{}                                & \multicolumn{1}{l|}{\textbf{Fixed Problems}} & \textbf{Pass rate} & \multicolumn{1}{l|}{\textbf{Fixed Problems}} & \textbf{Pass rate} & \multicolumn{1}{l|}{\textbf{Fixed Problems}} & \textbf{Pass rate} & \multicolumn{1}{c|}{}                                      & \multicolumn{1}{c}{}                                \\ \hline
Codellama-instruct-70b                                & \multicolumn{1}{l|}{589/1414}                & 41.65\%            & \multicolumn{1}{l|}{553/1401}                & 39.47\%            & \multicolumn{1}{l|}{728/1438}                & 50.62\%            & 43.96\%                                                    & \$4.20                                               \\ \hline
Phind-codellama-34b-v2                                & \multicolumn{1}{l|}{694/1414}                & 49.08\%            & \multicolumn{1}{l|}{550/1401}                & 39.25\%            & \multicolumn{1}{l|}{830/1438}                & 57.71\%            & 48.76\%                                                    & \$3.80                                               \\ \hline
WizardCoder\_instruct-33b                             & \multicolumn{1}{l|}{813/1414}                & 57.49\%            & \multicolumn{1}{l|}{708/1401}                & 50.53\%            & \multicolumn{1}{l|}{834/1438}                & 58.98\%            & 55.37\%                                                    & \$3.70                                               \\ \hline
Deepseek-coder-instruct-33b                           & \multicolumn{1}{l|}{893/1414}                & \textbf{63.15\%}   & \multicolumn{1}{l|}{971/1401}                & \textbf{69.30\%}   & \multicolumn{1}{l|}{971/1438}                & \textbf{64.81\%}   & \textbf{66.65\%}                                           & \$3.70                                               \\ \hline
Llama3-70b                                            & \multicolumn{1}{l|}{880/1414}                & 62.23\%            & \multicolumn{1}{l|}{755/1401}                & 53.89\%            & \multicolumn{1}{l|}{859/1438}                & 59.73\%            & 58.61\%\}                                                  & \$4.10                                               \\ \hline
\end{tabular}
\end{adjustbox}
\end{table*}

\paragraph{Experiment repetition}
Since there is some randomness in the response of LLMs, we repeated the experiment twice for all five models on the C++ dataset. The goal of the repetition is to verify the reliability of the results.

\section{Results}\label{sec:results}
\vspace{2mm}
\subsection{RQ1: Performance of Open-Source LLMs}
\begin{figure}[htb]
\vspace*{-0.5cm}
\centering
\includegraphics[scale=0.19]{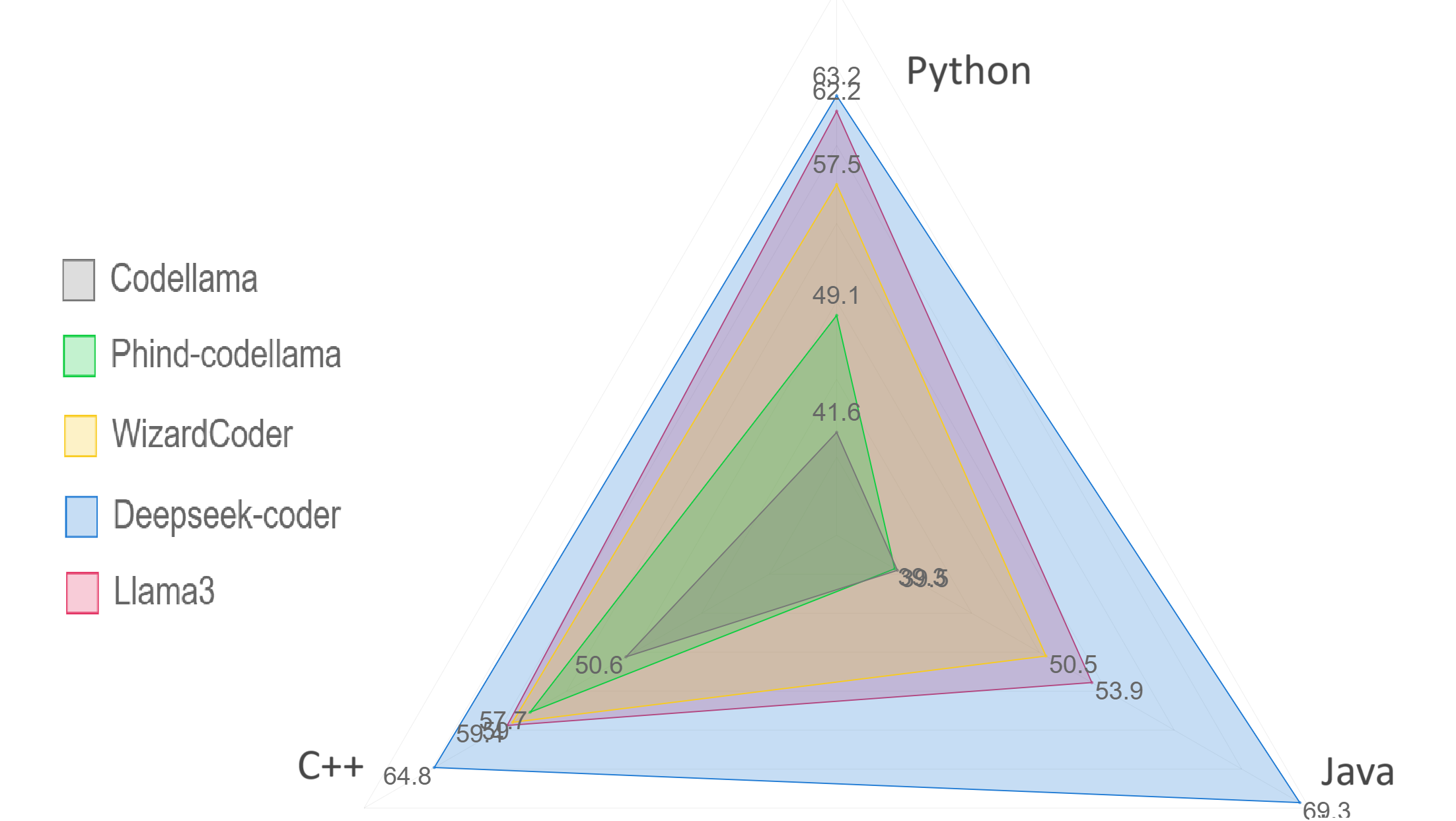}

\caption{A visualisation of the debugging performance of the LLMs for code in Python, C++ and Java}
\label{fig:radar}
\end{figure}

\vspace{1mm}
We report in Table \ref{tab:results_table},  the number of fixed problems for each model as well as the achieved pass rate for each programming language. 
We also report the average pass rate for all languages and  the costs of running the evaluation in USD. The evaluated LLMs achieved results ranging from 43.9\% to 66.6\%.  
The best score was achieved by the code model DeepSeek-Coder which  achieved a pass rate above 63\% for each of the languages. Both Codellama-instruct and Phind-Codellama achieved a pass rate below 50\%. Still, we notice that Phind-Codellama which is a fine-tuning of the Codellama-34B achieved better results than the larger Codellama-70B. Llama3, which is a general purpose LLM achieved almost 60\% pass rate, which is considerably better than Llama2-based code models.
In Figure~\ref{fig:radar} We see that DeepSeek-Coder  performed best on all three programming languages, while the Llama-2 based models (Codellama and Phind-Codellama) had the lowest scores for all three languages. This suggests that if a model is fluent in one programming language, then it is also likely to be fluent in other languages. 

Model size was not a determining factor in the performance of the models. In fact, some medium sized models, such as Phind-Codellama and DeepSeek-Coder, were able to achieve better scores than Codellama which had twice the size.

Regarding the costs of running the models, smaller models (33B and 34B) costed an average of 0.08 cent per code instance while larger models (70B) costed an average of 0.09 cent per code instance. So the difference in price between the models is negligible.

Closed source models, namely GPT-4 and GPT-3.5 from OpenAI achieved  75.0\% and 62.1\% respectively as reported by the DebugBench paper~\cite{tian2024debugbench}.
None of the evaluated tools was able to get a score that is comparable to GPT-4 and only DeepSeek-Coder was able to achieve results that are better than GPT-3.5

\vspace{3mm}

\noindent\fbox{%
    \parbox{\linewidth}{%
        \textbf{RQ1 answer:} There is a lot of variation between the scores of the different models, but some open source models were able to achieve decent results. Compared to top closed source models, only one open source was able to achieve results that are better than the GPT-3.5 score.
    }%
}
\subsection{RQ2: Relation between coding and debugging performance}
\begin{figure}[htb]
\centering
\includegraphics[scale=0.27]{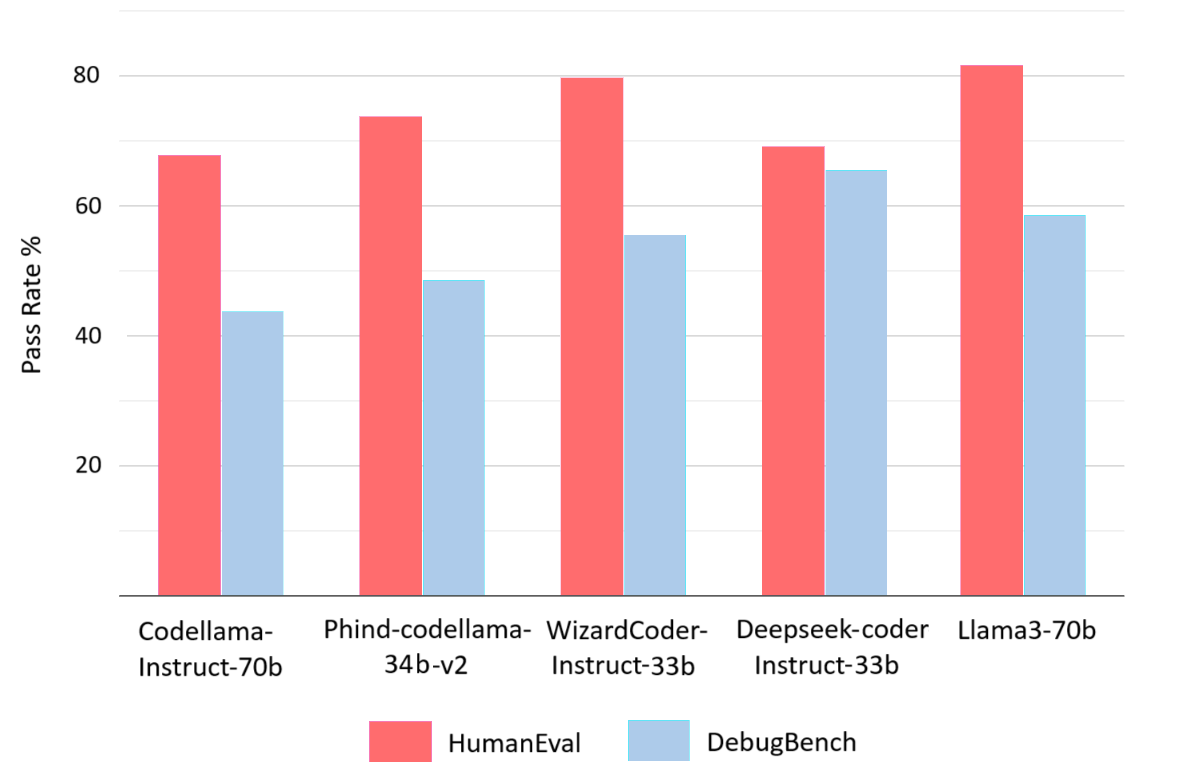}

\caption{Pass rates achieved by the open-source LLMs when evaluated on HumanEval and on DebugBench}
\label{fig:heval_dbench}
\end{figure}

We report the scores that the models achieved on HumanEval and on DebugBench in Figure~\ref{fig:heval_dbench}.
All five models achieved higher scores in the HumanEval coding benchmark compared to the benchmark DebugBench. In fact, all models were able to successfully solve the HumanEval coding problems with a pass rate ranging between 67\% and 81\%. The only benchmark that showed similar capabilities in coding and debugging is the DeepSeek-Coder model which achieved a pass rate of 69.2\% on HumanEval and a pass rate of 66.65\% on DebugBench.
When comparing the performance of the LLMs on both benchmarks, we see that except for DeepSeek-Coder, the models who had better results on HumanEval also achieved better scores in DebugBench.
From these observations, we see that although there is a hint that models that are better in coding might be better in debugging, we see no definite connection between  the performance of LLMs in coding, as evaluated by HumanEval, and their performance in debugging as evaluated by DebugBench.\\

\noindent\fbox{%
    \parbox{\linewidth}{%
        \textbf{RQ2 answer:} For four out of the five evaluated LLMs we noticed a relation between the performance of the models on HumanEval and their performance on DebugBench.  
    }%
}

\vspace{10pt}
\vspace{10pt}

\vspace*{-0.7cm}

\vspace{-2mm}
\section{Discussion}\label{sec:discussion}

\begin{table*}
\caption{Scores achieved for the C++ dataset during two runs and the calculated Mean and Standard deviation for each model }
\label{tab:2nd_cpp_results_table}
\begin{adjustbox}{max width=1\textwidth,center}
\renewcommand{\arraystretch}{1.5}%
\centering
\begin{tabular}{l|l|l|l|l}
\hline
\textbf{Model}  &  \textbf{First test pass rate}  & \textbf{Second test pass rate}  & \textbf{Mean} & 
\textbf{Standard deviation} \\ \hline
Codellama-Instruct-70b      & 50.62\%   & 51.55\%   & 51.08\%   & 0.465  \\ \hline
Phind-Codellama-34b-v2      & 57.71\%   & 55.94\%   & 56.82\%   & 0.885  \\ \hline
WizardCoder-Instruct-33b    & 58.98\%   & 55.79\%   & 57.38\%   & 1.595  \\ \hline
DeepSeek-Coder-Instruct-33b & 64.81\%   & 66.33\%   & 65.57\%   & 0.76   \\ \hline
Llama3-70b                  & 59.73\%   & 60.32\%   & 60.02\%   & 0.295  \\ \hline
\end{tabular}
\end{adjustbox}

\end{table*}

\subsection{Comparing with previous results}
There has been a few other works that evaluated the capabilities of LLMs for debugging. In this section, we compare our results with the results of other researchers.

First, we compare our results to the score reported by Tian et al~\cite{tian2024debugbench}, the authors of the DebugBench paper. The authors tested three open source models $Bloom$, $codellama$ and $codellama-inst$, and got a pass rate of 0.0 for all three of them. This means that none of the open source tools could repair any of the bugs. In our case, the models were able to achieve decent results.
This is probably due to the fact that all code models returned answers that contained not only code but also  explanatory text. The only model that returned code only was the general purpose llama3. The additional text makes the tests fail on LeetCode and this explains the 0.0 score obtained in the evaluation of Tian et al~\cite{tian2024debugbench}. In our experiment, we used a script to extract the code only from the answer and this lead to a positive performance of the models. 

Lee et al.~\cite{lee2024unified} compared the debugging performance of some closed source and open source LLMs using benchmarks in C, Java and Python. The authors generated 3 patches for each bug and looked for a plausible or correct patch among the generated responses. Codellama generated correct patches for 25/40 bugs in Java and for 33/40 bugs in Python, while DeepSeek-Coder achieved 30/40 and 25/40 correct patches for Java and Python respectively. The authors found that both GPT-3.5-Turbo-0125 (175B)  and GPT-4 generated a higher number of correct patches than the open source models. In our experiment, DeepSeek-Coder achieved better output  than codellama in all programming languages. This difference in the results might be due to the fact that Lee et al.~\cite{lee2024unified} used the DeepSeek-Coder-Base, while in our experiment we use DeepSeek-Coder-Instruct. The instruct version seems to have better coding capabilities as it is reported to achieve 69.2\% pass@1 on HumanEval compared to 50.3\% pass@1 for the base version.

In a study about vulnerability detection, Steenhoek et al.~\cite{steenhoek2024comprehensive} found that LLMs were unable to differentiate between buggy and fixed code. They report that LLMs performed only sligthly better than random guessing and that they performed far worse on complex debugging tasks from DBGBench. The evaluation was done using 100 functions from the SVEN dataset which is in C/C++ as well as the 27 bugs from DBGBench~\cite{bohme2017bug}. 
These low results could be explained by the complexity of the tasks performed in ~\cite{steenhoek2024comprehensive}. Previous research by  Huang and Changs~\cite{huang2023towards} has already shown that LLMs seem to be unable to manage complex tasks. The authors also note that existing benchmarks might be too simple to assess reasoning ability correctly~\cite{huang2023towards}. 



In another evaluation of code generation capabilities by Liu et al.~\cite{liu2024your}, the authors used HumanEval+ which is an improvement of the classic HumanEval. The authors found that the two open source models Phind-Codellama and WizardCoder achieved scores that are better than  ChatGPT but worse than GPT-4. In our experiment, both open source models achieved a score that is lower than GPT-4 and GPT-3.5. Since the authors did not specify the version of chatGPT  that was used, this could be due to the fact that they used a version of chatGPT that is less effective than GPT-3.5.


\vspace{-3mm}
\subsection{Contamination}~\label{sec:contamination}
One of the challenges in the evaluation of LLMs is contamination~\cite{sainz2023nlp,ravaut2024much,balloccu2024leak}. For example, Jain et al.~\cite{jain2024livecodebench} found that there was a drop in the performance of DeepSeek-Coder-Instruct on LeetCode problems that were released since September 2023, its release date. The authors interpret this as an indication of a potential contamination.
Although some research is being conducted on how to evaluate and remove contamination~\cite{yang2023rethinking}, decontamination does seem to be an easy task. In a study of code LLMs, Cao et al.~\cite{cao2024concerned} found that existing countermeasures for contamination such as using more recent data, using curated datasets or syntactic refactoring may not be effective.
Among the models we evaluated, only two mentioned using a decontamination strategy. OpenAI's decontamination methodology seems to have been applied to the Phind's dataset, and the  DeepSeek-Coder developers mention filtering out data from benchmarks such as HumanEval, MBPP, GSM8K and MATH. All other models didn't mention any decontamination strategy. 
The DebugBench data has been published on GitHub on January 9th, so all evaluated models had a knowledge cutoff date prior to the benchmark release. Although this might decrease the contamination threat, it  cannot fully eliminate it due to the fact that the LeetCode problems and solutions might have been used for pre-training the models.

\subsection{Result reliability}\label{sec:reliability}
LLMs are known to have randomness in their responses. So to check the reliability of the results, we run the experiment twice on the C++ dataset. We report the scores for both experiments in Table~\ref{tab:2nd_cpp_results_table}. We performed a statistical analysis of the results by calculating the mean and the standard deviation for each pair of measurements, the results are reported in the same Table. The deviations for all pair range between 0.29 and 1.59 which is relatively small. This indicates that the measurements are consistent and therefore likely reliable.

\section{Threats to Validity}\label{sec:threats}

\paragraph{Internal  validity}
Similarly to other LLM evaluations, we face a major internal threat due to the possible overlap between the training data and the evaluation dataset. We have already discussed possible contamination in Section~\ref{sec:contamination}. DebugBench was published after the knowledge cutoff date of the models. Although this might limit the threat, it cannot be fully eliminated because the used problems and their solutions existed on LeetCode before the benchmark release. The threat might also limited by the fact that the tests used to evaluate the debugging capabilities are not public and are only accessible for running via the LeetCode platform, so we know that these tests were not included in the training data of the evaluated LLMs.

The randomness in LLMs can also constitute a threat to the internal validity of the experiment. In fact, LLMs can produce different answers for the same prompt. To limit this threat, we repeated the experiment with C++ data twice. Our analysis in Section~\ref{sec:reliability} shows that the measurements are consistent and likely to be reliable.

\paragraph{External validity}
The main external validity threat lies in the benchmark code not being generalizable to other types of code. We argue that this threat is limited since the LeetCode dataset covers a variety of coding problems with different levels of difficulties. It also covers code in three different programming languages. Nevertheless, the results might not be generalizable to coding problems that are of different nature such as front-end  developement, or code that uses specific libraries.  In the future, We will evaluate the LLMs with more datasets.


\section{Related work}\label{relatedwork}

\subsection{Use of LLM for debugging}
The promising results about the capabilities of LLMs in software engineering lead to a surge in approaches that use LLMs to support debugging activities.
Kang et al.~\cite{Kang2023} introduced AutoSD, a method that leverages large language models and debuggers to automatically generate hypotheses and interact with buggy code, enabling conclusions to be drawn prior to patching. Feng et al.~\cite{feng2024prompting} presented AdbGPT, a lightweight approach that employs prompt engineering to reproduce bugs from reports automatically, without the need for training or hard-coding. Zhong et al.~\cite{Zhong2024} developed LDB, a debugging framework designed to assist large language models in refining generated programs by utilizing runtime execution data, segmenting programs into basic blocks, and tracking intermediate variable values after each block. Singh et al.~\cite{singh2024panda} proposed Panda, a framework aimed at providing context grounding to pre-trained large language models, thereby generating more useful and contextually relevant troubleshooting recommendations. Bouzenia et al.~\cite{bouzenia2024repairagent} introduced RepairAgent, an autonomous program repair agent that relies on a large language model. RepairAgent interleaves the processes of gathering information about the bug, collecting repair ingredients, and validating fixes, while dynamically deciding which tools to invoke based on the collected information and feedback from previous fix attempts.


\subsection{Evaluation of LLMs in debugging}
Several evaluations have been conducted to evaluate of the performance of LLMs in debugging. For example, Wu et al.~\cite{wu2023large} investigated the capabilities of ChatGPT-3.5 and ChatGPT-4 in fault localisation. Sobania et al.~\cite{sobania2023analysis} evaluated the bug fixing performance of chatGPT using the QuixBugs benchmark. Tian at al.~\cite{tian2024debugbench} evaluated the performance of five closed and open-source large language models using DebugBench.
Lee et al.~\cite{lee2024unified} compared their agent to other LLMs including two open-source LLMs, namely CodeLlama and DeepSeek-Coder.

Most of these works focused on evaluating the performance of the chatGPT, which is a closed source model. Only the works of Tian et al.~\cite{tian2024debugbench} and  Lee et al.~\cite{lee2024unified} cover some open source models. 
In this work, our goal was to evaluate different open-source LLMs and compare their capabilities.

\section{Conclusion and future work}\label{sec:con&FW}

In this work, we evaluated the debugging capabilities of five open-source large language models. The evaluation was done using DebugBench, a benchmark that includes a dataset of 4253 buggy code instances in Python, C++ and Java. Our results show that the capabilities of all the evaluated open-source LLMs are lower than the capabilities of the most recent closed-source model (GPT-4). Still, considering their relatively small size, some open-source LLMs were able to achieve decent results. For instance, DeepSeek-Coder which is only 33B in size achieved a score above 66\%.

One limitation of our evaluation, is that the used code instances are limited to one class only and are mostly  solutions to algorithmic problems. So In the future, we would like to evaluate open-source LLMs using  a wider variety of code types. Also we would like to evaluate the usefulness of the LLMs for practitioners when performing debugging tasks, with a special focus on complex tasks. Finally, we intend to explore how the debugging performance of open-source LLMs is impacted by prompt engineering and chain-of-thought prompting. 

\newpage
\bibliographystyle{unsrt}
\bibliography{ESEM}

\appendix

\end{document}